\documentclass[aps,pra,preprint]{revtex4-1}
\usepackage{graphicx,float}
\usepackage{amssymb}

\newcommand{\sgn}{\mathop{\mathrm{sgn}}}

\begin{document}
\title[Topological metrology]{Topological metrology and its application to optical position sensing}
\author{Nora Tischler$^{1,2}$, Mathieu L. Juan$^{1,2}$, Sukhwinder Singh$^{1,2}$, Xavier Zambrana-Puyalto$^{1,2,3}$,
Xavier Vidal$^{2}$, Gavin Brennen$^{1,2}$ and Gabriel Molina-Terriza$^{1,2}$}
\email{gabriel.molina-terriza@mq.edu.au}
\affiliation{$^1$Centre for Engineered Quantum Systems, Macquarie University, NSW 2109, Sydney, Australia}
\affiliation{$^2$Department of Physics \& Astronomy, Macquarie University, NSW 2109, Sydney, Australia}
\affiliation{$^3$Aix-Marseille Universit\'{e}, CNRS, Centrale Marseille, Institut Fresnel UMR 7249, 13013 Marseille, France}
\begin{abstract}
We motivate metrology schemes based on topological singularities as a way to build robustness against deformations of the system. In particular, we relate reference settings of metrological systems to topological singularities in the measurement outputs. As examples we discuss optical nano-position sensing (i) using a balanced photodetector and a quadrant photodetector, and (ii) a more general image based scheme. In both cases the reference setting is a scatterer position that corresponds to a topological singularity in an output space constructed from the scattered field intensity distributions.

\end{abstract}
\maketitle

\section{Introduction}

Measurements are the way to access physical properties of a system and are indispensable in natural sciences. However, the design and assessment of measurement strategies, which is the primary goal of the field of metrology, can be a complicated task e.g., due to the presence of noise and the intrinsic complexity of the system at hand. The \textit{precision} of a given metrology scheme can be assessed using metrics such as the Fisher information and the Cram\'er-Rao bound \cite{Kok}. The Fisher information determines the maximum amount of information about the parameter  that can be extracted from the system using a given measurement. The Cram\'er-Rao bound is an upper bound on the precision with which the parameter can be estimated and depends on the Fisher information. However, in practical experiments one is also interested in the \textit{robustness} of the given scheme to the presence of noise and in its \textit{versatility}, namely, its application to a broader set of systems.

In this paper we focus on the robustness of metrology schemes, and our analysis also naturally leads to a general metrology framework for designing versatile schemes. Inspired by the use of topological methods to classify physical phenomena both in the classical and quantum regimes \cite{Nakahara2003}, to enable fault tolerant quantum computing \cite{Nayak2008}, and for feature detection in image processing \cite{Freedman2009,Kalitzin1998,Kalitzin2001}, our strategy is based on identifying topological features from measurement outcomes. We exploit the inherent stability and robustness of these topological features \cite{Mermin1979}, which make the metrology scheme resilient to a large class of deformations, including stochastic noise and imperfections of the experimental method. We illustrate the approach in the context of nano-position sensing where we relate the reference position of the system to a \textit{topological singularity} apparent in the space constructed from measurement outcomes. We will present some examples where we demonstrate that the topological singularity used as a reference position of the system is only revealed through the measurement and data analysis, i.e. it is not physically attached to the system.  We also show that designing a metrology scheme based on topological features makes it robust to certain deformations \textit{of the physical system}, by showing how the topological singularity survives imperfections of the elements used in the positioning experiment or deformations of the target object. In general, we claim that designing a metrology scheme on topologically distinct features of the system leads to versatility and robustness of the method.

The paper is organized as follows. In Sec. \ref{Theory_section} we lay out the mathematical formalism and explain the general idea behind our topological approach. Section \ref{Applications_section} describes simple illustrations of our approach in the context of position sensing. We explain the familiar example of position sensing using a balanced photodetector, a quadrant photodetector, and then discuss a generalization of the image analysis at the output. Section \ref{Applications_section} may be read independently of Sec. \ref{Theory_section}. We conclude with a discussion in Sec. \ref{Discussion}. 

\begin{figure}[ht]
	\centering
		\includegraphics[width=6cm]{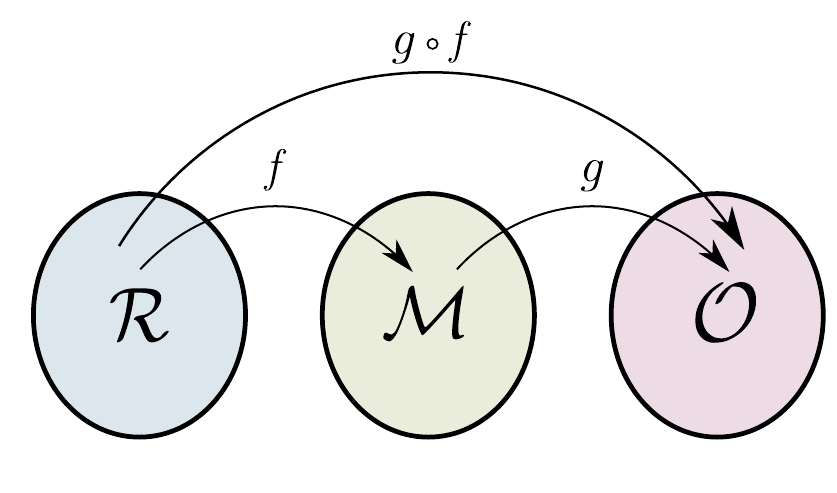}
	\caption{An abstract model of a metrology scheme consists of three topological spaces (i) $\mathcal{R}$ the space of physical parameters to be measured which describe the system, (ii) $\mathcal{M}$ the space of the outcomes of the measurements of the physical parameters, and (iii) $\mathcal{O}$ the ``order parameter'' space. The spaces are equipped with continuous maps $f:\mathcal{R} \rightarrow \mathcal{M}$ and $g:\mathcal{M}\rightarrow \mathcal{O}$. The desired metrological information, e.g. the reference position in a position sensing experiment, is extracted from the order parameter. A topological metrology scheme is one in which the metrological information is extracted from a \textit{topological feature}, such as a topological singularity, of the map $g \circ f$.} 
	\label{basic}
\end{figure}

\section{Topological metrology for reference determination}\label{Theory_section}

A general metrological setup typically consists of three components which can be modelled as topological spaces as shown in Fig.\ \ref{basic}: (i) $\mathcal{R}$ the space of parameters of the physical system, which are to be measured, (ii) $\mathcal{M}$ the space of measurement outcomes corresponding to those parameters, and (iii) $\mathcal{O}$ the ``order parameter'' space---the image of a map $g:\mathcal{M}\rightarrow \mathcal{O}$ that translates the bare measurement outcomes to a quantity that provides the desired metrological information. We also denote by $f:\mathcal{R} \rightarrow \mathcal{M}$ an abstract continuous function that maps, in a given trial of the metrology, the physical parameters to a measurement outcome.

In general, $\mathcal{R}$ could be the configuration space of positions and velocities for a classical dynamical system, or a Hilbert space describing a quantum mechanical system. To be more concrete here, we consider a generalized optical position sensing setup and in the following sections we will apply them to position sensing experiments. In this case $\mathcal{R}=\mathbb{R}^D$, the space of $D-$dimensional position vectors with coordinates $(x_1,x_2,\ldots, x_D)$ describing the location of the object. $\mathcal{M}$ is the space of  real scalar fields $I(y_1,y_2,\ldots,y_{D^\prime};x_1,x_2,\ldots, x_D)$ on $\mathbb{R}^{D^\prime}$, and these fields typically correspond to the measured intensity distributions. Notice that the dimensions ($D^\prime$) of the space where the intensity measurements are taken do not need to be the same as the dimensions ($D$) of the space of physical positions to be determined. In practice, $\mathbb{R}^{D^\prime}$ is discretized to the set of points $\{i_1, \cdots,i_{D^\prime}\}$, where each index $i_k$ can take values from $i_k=1,\ldots,n_k$, with $\prod_{k}n_k =N$, where N would represent the total number of image pixels, such that the continuous function is mapped to a discrete set: $I(y_1,y_2,\ldots,y_{D^\prime}; x_1,x_2,\ldots, x_D)\rightarrow \{I_{i_1,i_2,\ldots i_{D^\prime}}(x_1,x_2,\ldots, x_D)\}$. We will focus on the particular goal of determining a \textit{reference position} in the experiment, which may be required for example to align different components of the system or to reliably return to an initial position of the system. We require that the method should still yield a reference position in the presence of noise, such as intensity fluctuations of the input optical field, or electronic noise in the detectors, and possibly other deformations in the experiment, such as differently shaped target objects or differences in the efficiency and gain of the detectors. Here we advocate to introduce an order parameter space and the associated map $g$ such that the reference position corresponds to a \textit{topological singularity} in the composition $g \circ f$. To concretely demonstrate the topological analysis, we choose our order parameter space $\mathcal{O}$ to be the space of real unit vectors $v=v(x_1,x_2,\ldots, x_D)$ in the $D$ dimensional Euclidean space $\mathbb{E}^D$ with components denoted  $v_j$. All vectors at non-singular points are normalised as $\sum_j v_j^2=1$. From now on, summation over repeated indices is assumed. We have chosen that the number of components of the vector $v$ is the same as the number of dimensions in our reference space, $D$, because in that case the only possible stable topological singularities that can appear will have dimension zero (points) \cite{Mermin1979}, which matches our aim to find a reference position. Let $V$ be some region in $\mathbb{E}^D$ and $S$ its closed boundary: $S=\partial V$.  At the non-singular points of $v$, define the differential $D-1$ form 
\begin{equation}\label{omegaEqn}
\omega=v_{j_1} dv_{j_2}\wedge dv_{j_3}\wedge\cdots\wedge dv_{j_D}\epsilon^{j_1j_2\cdots j_D}
\end{equation}
where  $\epsilon$ is the rank $D$ totally anti-symmetric tensor.  Consider the following quantity
\[
\mu=\frac{\Gamma(D/2)}{2\pi^{D/2}}\oint_{S} \omega,
\]
with $\Gamma(x)$ being the gamma function. Appendix \ref{app_topol} contains a formal derivation of the following important facts: The quantity $\mu$ can only have an integer value and is a topological invariant of the system. The actual value of $\mu$ can be calculated by summing the topological indices of singular points, $p$, of the system, by deforming the contour of integration, i.e. $\mu=\sum_p \mu_p$. Here $\mu_p$ can be formally defined as
\begin{equation} \label{mupEqn}
\mu_p=\frac{\Gamma(D/2)}{2\pi^{D/2}}\oint_{S_p} \omega\in\mathbb{Z}-\{0\},
\end{equation}
where $\oint_{S_p}$ surrounds a neighbourhood $V_p$ small enough that it only contains singularity $p$.

The following are some illustrative examples of the use of this formalism.
\begin{enumerate}
\item
$D=1$. A system whose position is mapped to an intensity field $I_{i_1}(x_1)$, with $i_1=\{1,2\}$. Here, we have chosen $D^\prime=1$ and discretized this space so that $N=2$. An order parameter which is sensitive to roots in the intensity function is $\vec{v}(x_1)=g \circ f (x_1) = \sgn(I_2(x_1)-I_1(x_1))\hat{y}_1$. Over the closed interval $V=[a,b]\in \mathbb{R}$ with boundary $S=\{a\}\cup \{b\}$, the index is
\[
\mu=\frac{1}{2}(g \circ f(b)-g \circ f(a)).
\]
This value will not change under deformations of the interval $S$ provided a zero of the intensity is not crossed. 
\item
$D=2$ and $D^\prime=0$, so that $N=1$. A system whose position is mapped to an intensity field $I(x_1,x_2)$. An order parameter which is sensitive to critical points in the intensity field is the normalised gradient $\vec{v}(x_1,x_2)=g \circ f (x_1,x_2)=\frac{\vec{\nabla} I(x_1,x_2)}{(\partial_{j} I(x_1,x_2) \partial_{j} I(x_1,x_2))^{1/2}}$. The invariant 
\[
\begin{array}{lll}
\mu&=&\frac{1}{2\pi}\oint_{S}(v_1dv_2-v_2dv_1)\\
&=&\frac{1}{2\pi}\oint_S \Bigg[\frac{\partial_1 I}{(\partial_j I \partial_j I)^{1/2}} \Big(\frac{d\partial_2 I}{(\partial_j I \partial_j I)^{1/2}}-\frac{\partial_2 I (\partial_j I d\partial_j I)}{(\partial_j I\partial_j I)^{3/2}}\Big)\\
&&-\frac{\partial_2 I}{(\partial_j I \partial_j I)^{1/2}} \Big(\frac{d\partial_1 I}{(\partial_j I \partial_j I)^{1/2}}-\frac{\partial_1 I (\partial_j I d\partial_j I)}{(\partial_j I\partial_j I)^{3/2}}\Big)\Bigg]\\
&=&\frac{1}{2\pi}\oint_S \frac{\partial_1 I d\partial_2 I-\partial_2 I d\partial_1 I}{(\partial_j I \partial_j I)}.
\end{array}
\]   
Here the index is none other than the winding number. For example, if $I(x_1,x_2)=ax_1^2-bx_2^2$ for $a,b>0$ then choosing $S$ to be the unit circle centred at the origin, $\mu_{p=(0,0)}=-1$.
\end{enumerate}

Having established the general criteria for a topological invariant determined by the unit vector field, we now cast the problem of measuring physical observables in that setting. The key observation is that we can perform measurements on a physical system and construct a unit vector field from those measurements in a way that singular points provide useful information, such as the centre of mass position or spatial symmetry of an object.  In many circumstances, direct detection of a singular point may be difficult due to finite resolution detectors, noise in the measurement, etc.  However, indirect detection is always possible by computing the index $\mu_p$ over a closed path surrounding a region in which the point is proposed to exist, by Eq.\ (\ref{mupEqn}).     

We wish to characterise the degree of stability of the index $\mu_p$ to noise and finite resolution detection.  As stated, the composition $g\circ f$ defines a mapping from $\mathcal{R}\rightarrow \mathcal{O}$. The quantity $\mu_p$ is invariant to continuous changes in $f$ and $g$ which deform the region $V_p$ but do not introduce new singularities that cross $S_p$. Also, $\mu_p$ is invariant under certain rescalings.  For example, write the mapping $g$ to the order parameter as a composition $g=r\circ\tilde{g}$ where the image of $\tilde{g}$ is the unnormalised vector field: $\tilde{v}(x_1,\ldots, x_d)=\tilde{g}(I)$, and $v=r(\tilde{v})$.  Then arbitrary rescalings $\tilde{v}_j\rightarrow \lambda_j\tilde{v}_j$, where $\lambda_j>0$, will leave $\omega$ (Eq.\ (\ref{omegaEqn})) and hence $\mu_p$ (Eq.\ (\ref{mupEqn})) invariant. In fact if one only cares to detect the presence of a singularity, then the sign of $\mu_p$ is irrelevant and it suffices to have $\lambda_j$ all strictly positive or all strictly negative everywhere.

\section{Application to position sensing}\label{Applications_section}
\subsection{Physical setup}

\begin{figure}[ht]
	\centering
		\includegraphics[width=15cm]{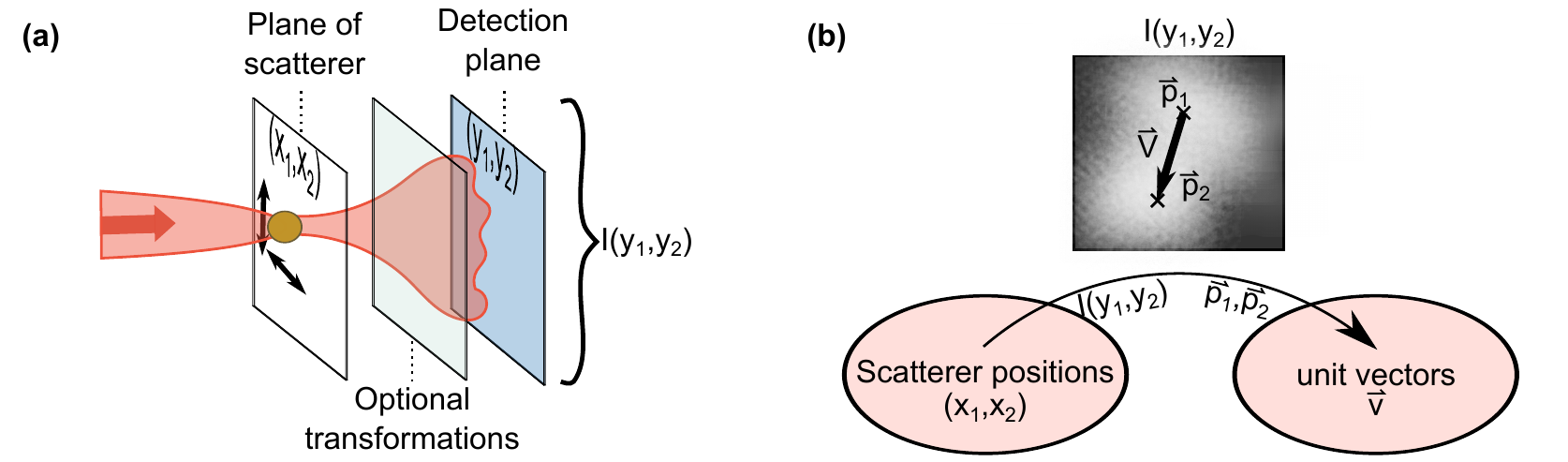}
	\caption{(a) Physical system for optical position sensing: An incident light field scatters off the object in question, undergoes optional further transformations, and is measured in a plane behind the scatterer. The aim is to identify one specific position of the scatterer, which can be translated transversely, based on the measured intensity distribution. (b) Mapping from the $\mathbb{R}^2$ space of scatterer positions (reference space $\mathcal{R}$) to unit vectors (in the order parameter space  $\mathcal{O}$) through the image analysis of the intensity distribution in space $\mathcal{M}$. The intensity image corresponds to a single scatterer position $(x_1,x_2)$, and thus we have dropped the dependence on these variables for the intensity, i.e. $I(y_1,y_2;x_1,x_2)\equiv I(y_1,y_2)$} 
	\label{Physical_system}
\end{figure}

We now describe position sensing as a concrete example for a metrology scheme where the reference position is determined from a topological singularity. The physical system consists of an optical illumination field, a scatterer (e.g., a nanoparticle or a nanohole) that can be translated in the transverse plane, and a detection plane, as illustrated in Fig.\ \ref{Physical_system} (a). The scatterer position is a vector in $\mathbb{R}^2$, which corresponds to the space $\mathcal{R}$ of the previous section. The optical intensity distribution is measured at the detection plane. Then, the space $\mathcal{M}$ corresponds to all possible scalar intensity distributions in $\mathbb{R}^2$. Given that the scatterer can be moved at will, the aim is to reliably identify a reference position that we can easily return to. This will allow us to measure displacements from that position or simply use it to stabilize the position of the scatterer. We then seek a reference position whose existence is robust to noise in the measurement.
As we are in the linear optics regime, a small change in the illuminating field or in the shape or properties of the scatterer will result in a small change in the output field. This means that if we can find an idealized situation where the metrological scheme presents a topological singularity in the order parameter space, the structural stability theorems in the previous section will ensure that the topological singularity is also present when there are small deviations from the ideal situation.


\begin{figure}[ht]
	\centering
		\includegraphics[width=8.5cm]{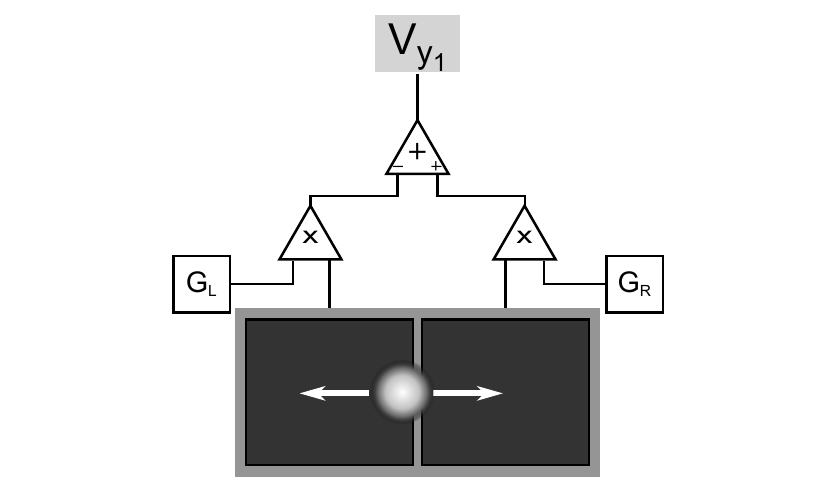}
	\caption{The balanced photodetector, a 1 dimensional position sensing device. The power incident on each of the two detector regions is amplified by a gain, and then the difference between the signals is evaluated.} 
	\label{BPD.fig}
\end{figure}

\subsection{Example in 1 dimension: Balanced photodetector}\label{BPD}
A common position sensing device used to measure the position of a scatterer along one axis is the balanced photodetector, illustrated in Fig.\ \ref{BPD.fig}. One property that makes this method particularly popular is that it is quite robust under the effect of noise and other imperfections of the system. Such robustness was already apparent in the predecessor of position sensing devices \cite{Wallmark1957}. We will now show that this method can be analyzed with the tools explained in the previous section. This will unveil that the main properties of this metrological method are founded on the identification of a topological singularity of the vector field constructed from the measurements. 

A balanced photodetector consists of two adjacent photodiodes with gain amplification factors $G_L$ and $G_R$. For each scatterer position, we take a measurement of the power incident on the two regions, $P_{i}$, $i\in\{\mathrm{L},\mathrm{R}\}$ which is amplified to produce output signals $S_{i}=P_{i}\times G_{i}$. A differential signal, $\Delta S=S_{\mathrm{R}}-S_{\mathrm{L}}$, is calculated from each measurement. The $x_1$-coordinate of the scatterer at which the signals are balanced, that is $\Delta S=0$, can be used as a reference. In order to formally follow the analysis in Sec. \ref{Theory_section}, we will first consider the differential signal as a vector in $\mathbb{R}$: $\vec{V}=\Delta S \hat{y}_1$. Then, we build the normalized vectors ($\vec{v}=\vec{V}/|\vec{V}|=\sgn(\Delta S)\hat{y}_1$) in order to carry out the analysis in the order parameter. Now, the equivalence to the 1 dimensional case discussed earlier  becomes evident. The map $g\circ f$ simply maps the $x_1$-coordinate of the scatterer to a unit vector in $\mathbb{R}$ and the position where the scatterer balances the detector ($\Delta S = 0$), appears as a topological singularity in the order parameter.

\begin{figure}[t]
	\centering
		\includegraphics[width=12cm]{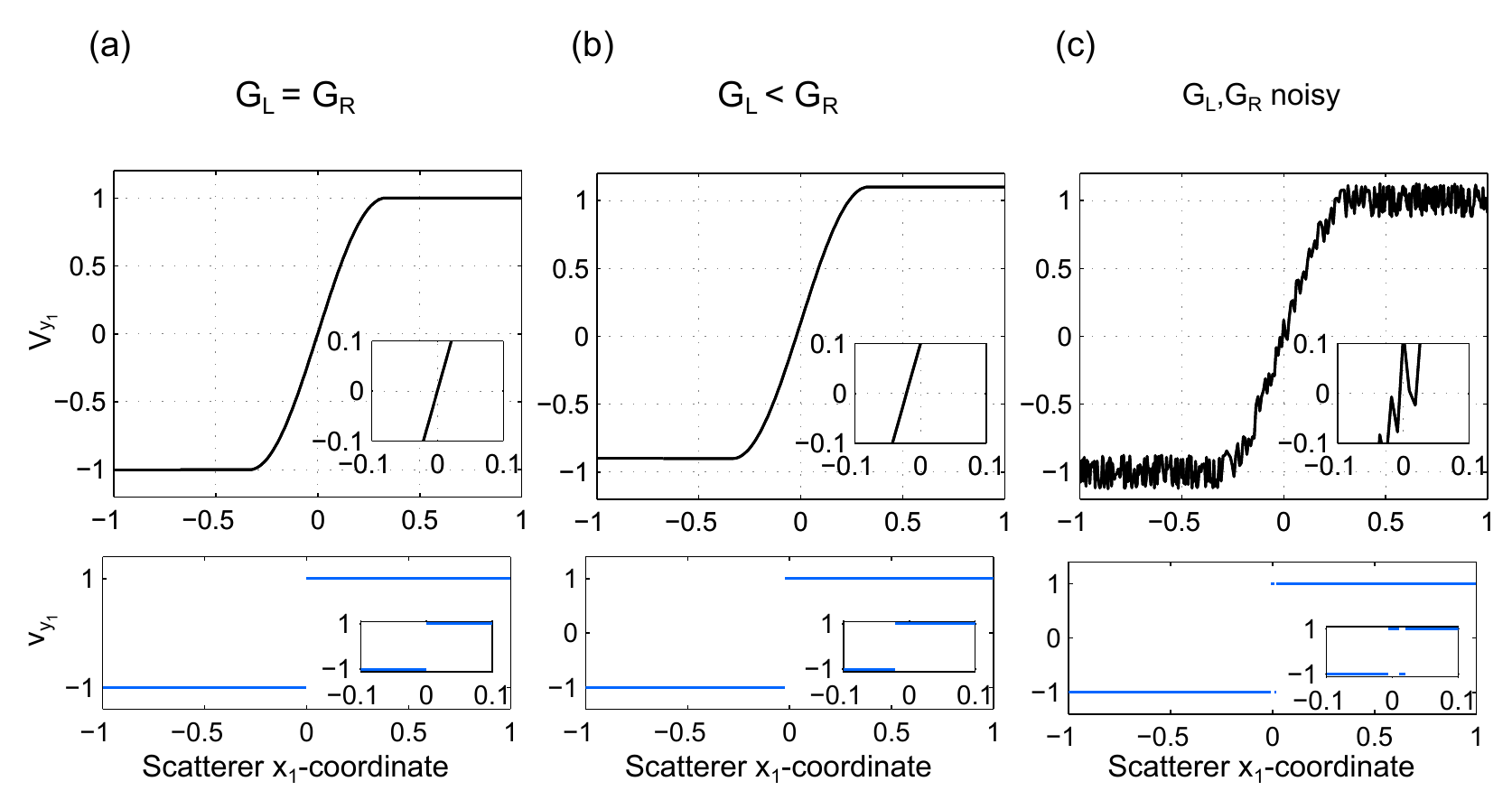}
	\caption{Typical output values $V_{y_1}$ and $v_{y_1}$ from a balanced photodetector with respect to the position of the scatterer in the $x_1$-coordinate for (a) an ideal physical system, (b) unequal gains, and (c) equal gains but with added noise.} 
	\label{Kinks}
\end{figure}

Figure \ref{Kinks} illustrates the effect of deformations of the system on the measurement results. Shown are the only component of the unnormalised $\vec{V}$ and normalised $\vec{v}$ as a function of scatterer $x_1$-coordinate for three different cases: (a) a case where the gains of the two regions are equal and no noise is present, (b) a case where the gains are unequal and still no noise is present, and (c) a case with equal gains but in the presence of stochastic noise with a bounded amplitude. Here we have considered the power gains as a perturbation of the physical system. Another component that could change is the probe light field. Since case (a) corresponds to an ideal physical system, its result can be regarded as the unperturbed reference position. The imbalance of the gains in (b) results in a perturbation that maintains the monotonicity of $V_{y_1}$ but shifts the location of the single zero crossing---the singularity in $v_{y_1}$. The noise in (c) breaks the monotonicity of $V_{y_1}$ and introduces further crossings in a way that the topological index $\mu$ remains invariant, which in the 1 dimensional case can be -1,0, or 1, depending on the sign of $v_{y_1}$ at the boundaries of the interval in question (here $\mu=1$). Since the maximum amplitude of the noise is limited, the new crossings in $V_{y_1}$ and hence singularities in $v_{y_1}$, lie within a bounded distance from the unperturbed reference location (corresponding to case (a)). In this case, it is no longer possible to pinpoint a single singularity as the reference, but we can still identify an interval that must contain the reference. This illustrates that while noise can impact the precision of the measurement, the existence of the reference position can nonetheless remain robust. This simple example should make our motivation for relating the reference position to a topological singularity apparent.

\subsection{Example in 2 dimensions: Quadrant photo detector}

The quadrant photo detector (QPD) provides a simple extension of the position sensing scheme just described to 2 dimensions. A QPD consists of four detector regions with amplification gains $G_{i}$, $i\in\{\mathrm{NE},\mathrm{NW},\mathrm{SW},\mathrm{SE}\}$, and arranged as the four quadrants of the measurement plane. 
We perform a measurement of the power incident on the quadrants, $P_{i}$ which is amplified to produce output signals: $S_{i}=P_{i}\times G_{i}$. Two differential signals are calculated from each measurement, one horizontal difference of the signals, $\Delta S_{\mathrm{h}}=S_{\mathrm{NE}}+S_{\mathrm{SE}}-S_{\mathrm{NW}}-S_{\mathrm{SW}}$, and one vertical difference, $\Delta S_{\mathrm{v}}=S_{\mathrm{NE}}+S_{\mathrm{NW}}-S_{\mathrm{SE}}-S_{\mathrm{SW}}$. The scatterer position at which the signals from the four quadrants are balanced can be used as a reference point, which is equivalent to the condition $\Delta S_{\mathrm{h}}=\Delta S_{\mathrm{v}}=0$. This case is similar to the 2 dimensional case discussed in Sec. \ref{Theory_section} after we identify $\Delta S_{\mathrm{h}}$ and $\Delta S_{\mathrm{v}}$ as components of a vector $\vec{V}=\Delta S_\mathrm{h} \hat{y}_1 + \Delta S_\mathrm{v} \hat{y}_2$ in $\mathbb{R}^2$, which is normalised to obtain $\vec{v} = \vec{V}/\sqrt{\vec{V}\cdot\vec{V}}$. That is, the condition $\Delta S_{\mathrm{h}}=\Delta S_{\mathrm{v}}=0$ corresponds to a topological singularity in the vector field $\vec{v}$. In the idealised case where the gain factors of the quadrants are equal and both the scatterer and probe light have discrete four fold rotational symmetry (symmetry under rotations by $\pi/2$ about an axis through the origin perpendicular to the scatterer plane), it is clear that both differential signals are zero when the scatterer is placed at the symmetric position. On the other hand, this position corresponds to a singularity of the order parameter, as argued in Sec. \ref{Theory_section}. In fact, such a reference position exists even when the light field and scatterer become asymmetric, and when the gain factors become unequal, since they correspond to continuous deformations on the order parameter space. This effect is equivalent to the one shown for the one dimensional case. The topological singularity is displaced due to the deformations in the gain or probe beam, but its structural stability avoids its destruction, unless the parameters change so radically that the symmetry of the system is completely lost. Again, a proper analysis shows that the QPD being robust, versatile, and easy to operate, which contributes to its popularity, stems from the topological features of the method. 

\subsection{Image-based nanopositioning using a topological singularity}
Finally, we describe a new, more general image-based position sensing approach, within the same physical setup described above, where the reference position is once again obtained from a topological singularity in the order parameter (space $\mathcal{O}$) that is constructed from the output images. In a measured intensity distribution we define two points $\vec{p_1}$ and $\vec{p_2}$ and consider the difference vector, $\vec{V}=\vec{p}_2-\vec{p}_1$, normalized to a unit vector $\vec{v} = \vec{V}/\sqrt{\vec{V}\cdot\vec{V}}$ in $\mathbb{R}^2$. In this way, we map each position of the scatterer to a unit vector $\vec{v}$ in $\mathbb{R}^2$, see Fig.\ \ref{Physical_system} (b). The vector field $\vec{v}$ corresponds to our order parameter as a function of the configuration space position. Our goal is to obtain a vector field $\vec{v}$ that changes continuously with the scatterer position, and that exhibits a topological singularity which can be used as a reference position.

How can we choose the two points $\vec{p}_1$ and $\vec{p}_2$ to generate such a vector field? We can always keep one of the points, say $\vec{p}_1$, fixed. Then as long as $\vec{p}_2$ shows a continuous dependence on the scatterer position, the vector field also varies continuously with the scatterer position. A quick analysis will show that this is a straight generalization from the BPD or QPD systems to an arbitrary large number of detectors, or pixels of a camera. Another, more interesting, option is to have both points depend (differently) on the intensity distribution. Care has to be taken when choosing the points to fulfil the conditions expressed in Sec. \ref{Theory_section}. For instance, one could consider using the global minimum of the intensity distribution as one of the points. However, even when the intensity varies continuously with the scatterer position, the global minimum may undergo a discrete jump (e.g. the global minimum could jump from one trough in the intensity distribution to another distant trough that has approximately the same depth, even when the scatterer position is changed slightly). Such a choice of image analysis does not guarantee the desired property that the vector field changes continuously under a continuous change of the parameters.

Here we consider defining the points $\vec{p}_i$ ($i=1,2$) as weighted centroids, which satisfy the stated continuity requirement:
\begin{equation}\label{eq:centeroid}
\vec{p}_i=\int \mathrm{h_i}(I(y_1,y_2;x_1,x_2))(y_1\hat{y}_1+y_2\hat{y}_2)\mathrm{d}y_1\mathrm{d}y_2
\end{equation}
where $\mathrm{h_i}(I(y_1,y_2;x_1,x_2))$ is any continuous function of the intensity value at the coordinate $(y_1,y_2)$ in the image, when the scatterer position is $(x_1,x_2)$. If the image is discretized due to uniform rectangular pixelation, Eq. (\ref{eq:centeroid}) is replaced with
\begin{equation}
\vec{p}_{i}=\sum_j \mathrm{h_i}(P_j)(\bar{y_1}_j\hat{y}_1+\bar{y_2}_j\hat{y}_2), 
\end{equation}
where $j$ labels the pixels, $P_j=\int_{A_j} I(y_1,y_2;x_1,x_2)\mathrm{d}y_1\mathrm{d}y_2$ with $A_j$ being the area of pixel $j$, and $\bar{y_1}_j$, $\bar{y_2}_j$ are the mean $y_1,y_2$ positions of pixel $j$. 

\begin{figure}[h!]
	\centering
		\includegraphics[width=8.5cm]{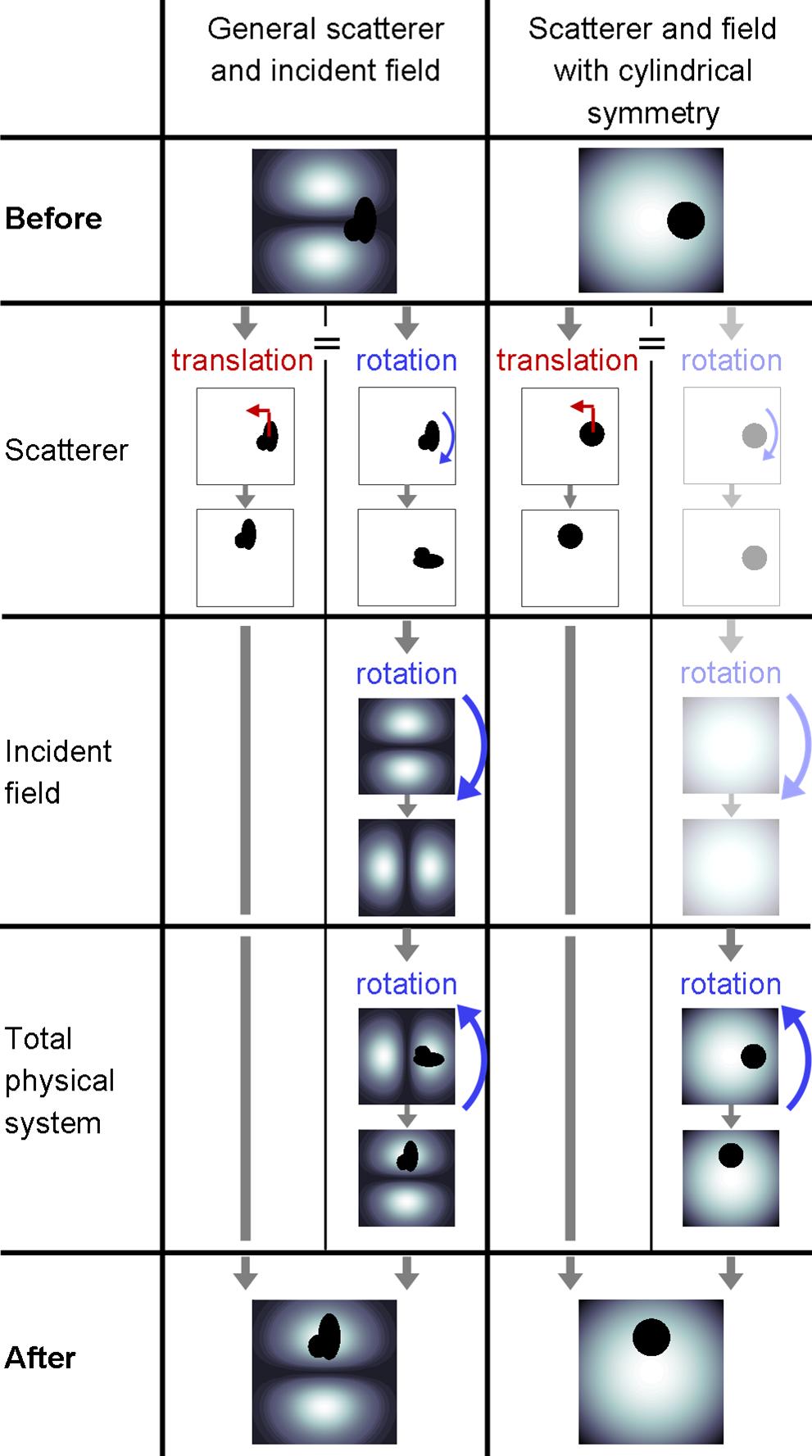}
	\caption{Relationship between two optical systems where the scatterer has been translated. Left column: For a general scatterer and incident field, the translation of the scatterer can be replaced by a rotation of both incident field and scatterer, followed by a rotation of the whole physical system in the opposite direction. Right column: When the scatterer and incident field are both cylindrically symmetric and translation occurs between two positions of equal distance from the center of the incident beam, the same relationship applies. However, the rotations of the scatterer and incident field about their respective axes have no effect. Subsequent rotation of the complete system results in rotation of the scattered field pattern.} 
	\label{Symmetry_argument}
\end{figure}

When considering a cylindrically symmetric light field and scatterer, the intensity distribution has a rotational symmetry. As shown in the appendix, when the position of the scatterer $(x_1,x_2)$ is such that it lies on the symmetry axis, the cylindrical symmetry of the system enforces the weighted centroids in Eq. (\ref{eq:centeroid}) to be zero, and the vector $\vec{v}$ in the order parameter has a discontinuity, $\vec{p}_1=\vec{p}_2=\vec{0}$, provided a coordinate change in the image space to have the origin coincide with the symmetry axis. Thus, one can conveniently find a vector discontinuity in $\mathcal{O}$ in the presence of rotational symmetry. But what guarantees in this case that this vector discontinuity is indeed a topological singularity? By symmetry, a translation of the scatterer between two positions equally distant from the center is equivalent to a rotation of the complete physical system, as illustrated in Fig.\ \ref{Symmetry_argument}. This in turn leads to a rotation of the intensity distribution and of the vectors $\vec{p}_{1}$ and $\vec{p}_{2}$. On the other hand, when the scatterer is placed at the central location, the two points coincide, resulting in a zero vector. Now, to prove that this scatterer position gives rise to a point singularity in the order space, choose a new scatterer position $\vec{x}$ that lies in the neighbourhood of this central position. Due to the symmetry breaking, the vectors $\vec{p}_{1}$ and $\vec{p}_{2}$ will no longer coincide, and the order parameter vector $\vec{v}$ will be well defined. As the scatterer traverses a concentric circle, $\vec{v}$ undergoes one complete rotation through $2\pi$. Thus, the central point is a singularity with winding number $\mu=1$.

In a real experiment, perturbations generally break the cylindrical symmetry of the system. However, the same arguments used in the previous cases hold, and the topological singularity is robust under noise or perturbations of the system. Then, in the experiment, the topological singularity, which is related to the reference position, still exists as long as the corresponding perturbations in the vector field are continuous and sufficiently small. This, in fact, is the main motivation for relating the reference position to a topological feature obtained from the measurement outcomes.

An interesting feature of this choice of image analysis is that it renders this positioning technology self-referenced. The reference position does not depend on the external devices used to measure it, as was the case with the QPD and BPD. In this case, the reference position only depends on the incident beam and the scatterer. In the perfectly cylindrically symmetric situation it will coincide with the axis of symmetry of the incident beam and the scatterer. In real experimental situations it will correspond to the position of the stable topological singularity of the system.

\section{Discussion}\label{Discussion}

In this work we have presented a new way of analyzing metrological systems based on the geometrical properties of the experimental set-up and posterior analysis. These geometrical properties can be unveiled with the use of mathematical techniques in the field of topology. Then, the distinct features of the metrological system can appear as topological singularities in an order parameter derived from the measurement. We have shown that when such an approach is possible, the topological singularity serves as a distinct point in the system, which is robust under continuous perturbations. This can be used to set a reference point.

Topological singularities present a structural stability under continuous changes of parameters. Consequently, the existence of a reference point linked to that topological singularity will be particularly resilient to both external noise and the influence of small changes of parameters of the metrological set-up. We have exemplified all these features in the context of position metrology. Using the proposed approach, we have analyzed two common metrological systems used to provide a reference point (the balanced photodetector and the quadrant photodetector), and showed that most of the interesting features of the methods (including robustness and versatility) are connected to the fact that they are referenced to a topological singularity of the vector field obtained through measurement and data analysis. We have also proposed a new topological metrology method for position sensing, which offers an additional advantage that it does not rely on an external reference point.

Topological concepts have already been used in image processing algorithms \cite{Kalitzin1998,Kalitzin2001}. An important distinction is that the topological singularities in those references are at a different level compared to what was discussed in this paper. In the references, the vector field is obtained from a single image, and a topological singularity is identified in the image plane (the equivalent of $(y_1,y_2)$ in Fig.\ \ref{Physical_system} (a)). In contrast, we are taking an image for each coordinate in the reference space, which is the scatterer position $(x_1,x_2)$. We do not seek to identify a point in the image plane as the reference point; each image only provides one vector of the vector field. The topological singularity then resides in the vector field that is constructed in the analysis using multiple images. In this way, a point in the configuration space, i.e., a particular reference position of the scatterer, can be identified as the reference. We make this distinction by writing that our topological singularities are in an order parameter, constructed from measurement outcomes, as a function of the configuration space setting. We would like to stress that this concept is not limited to image-based measurements. The formalism presented is very general and could accommodate other metrological methods, such as measurements of magnetic fields, etc. The key ingredients are that the measurements have to be relative measurements, as most measurements are, and thus a reference point must be found, and that such reference point can be linked to a geometrically distinct point of the metrological system, in a way that when analyzed appropriately, it can give rise to a topological singularity in the order parameter space.

As we have shown, the practical use of topological singularities in a vector field constructed from measurement outcomes is key to understanding already existing metrological systems. The identification and analysis of topological features allows us to understand what makes such techniques and devices robust. Moreover, the analysis could also be applied to the design of new metrology schemes. An important point for further study is that from the analysis performed in this work, there is no immediate connection to the measurement precision. Consequently, optimal precision is not guaranteed by designing a technique in the way we have put forward. However, assuming that the precision of the measurement is limited by the precision in the location of the reference point, one could in principle use the theorems on the structural stability of the topological dislocations in order to give upper bounds to the precision of a topological metrology system for a given noise spectrum. On the other hand, we have discussed the robustness of the system. This is of practical importance because it affects the versatility of the method: Methods that use a topological singularity of an order parameter constructed from measurement outcomes as a reference are expected to be robust to deformations, and therefore, applicable to a number of physical systems.
\section{Acknowledgements}
This work was funded by the Center of Excellence in Engineered Quantum Systems (EQuS). G.M.-T is also funded by the Future Fellowship FF110100924.

\appendix
\section{Topological invariants}\label{app_topol}
The contents of this appendix can be found in most elementary textbooks in topology and other specialized texts \cite{Nakahara2003,Mermin1979}. We repeat it here for the convenience of the non-specialist reader. Let us define again, for the sake of completeness, the two relevant spaces:  $\mathcal{R}=\mathbb{R}^D$ and the order parameter space $\mathcal{O}$ as the space of real unit vectors $v=v(x_1,x_2,\ldots, x_D)$ in the $D$ dimensional Euclidean space $\mathbb{E}^D$ with components denoted  $v_j$. At the non-singular points of $v$, we also define the differential $D-1$ form 
\begin{equation}
\omega=v_{j_1} dv_{j_2}\wedge dv_{j_3}\wedge\cdots\wedge dv_{j_D}\epsilon^{j_1j_2\cdots j_D}
\end{equation}
where  $\epsilon$ is the rank $D$ totally anti-symmetric tensor. Finally, we define the topological index of the volume $V$ as the quantity 
\[
\mu=\frac{\Gamma(D/2)}{2\pi^{D/2}}\oint_{S} \omega
\]
where it is assumed that $\omega$ is defined everywhere on the surface $S$ surrounding the volume $V$. The normalisation comes from the formula for the surface area of $S^{(D-1)}$ the unit sphere in $D-1$ dimensions, with $\Gamma(x)$ being the gamma function. The quantity $\mu$ is a topological invariant of the volume enclosed by surface $S$.

Because of the normalisation condition,  $d(v_jv_j)=2v_jdv_j=0$, at any non-singular point there is some non zero $v_k$ such that $dv_k=\frac{1}{v_k}\sum_{j\neq k}v_jdv_j$. Hence at non-singular points, 
\[
\begin{array}{lll}
d\omega&=&dv_{j_1} \wedge dv_{j_2}\wedge dv_{j_3}\wedge\cdots\wedge dv_{j_D}\epsilon^{j_1j_2\cdots j_D}\\
&=&D! dv_{j_1} \wedge dv_{j_2}\wedge dv_{j_3}\wedge\cdots\wedge dv_{j_D}\\
&=&0.
\end{array}
\]
The first line follows because  $dd=0$, the second from the antisymmetry of the wedge product, and the third from the aforementioned fact that one of the components $dv_k$ is not independent and $dv_j \wedge dv_j=0$.
If the region $V$ contains no singular points, then $\omega$ is defined everywhere in $V$ and from the generalised Stoke's theorem
\[
\mu=\frac{\Gamma(D/2)}{2\pi^{D/2}}\oint_{S} \omega=\frac{\Gamma(D/2)}{2\pi^{D/2}}\oint_{V} d\omega=0.
\]
However, when the region does contain one or more singular points, the form $\omega$ (from Eq.\ \ref{omegaEqn}) is not defined everywhere in the region and we must be more careful.  Let us focus on some singular point $p$ and denote its neighbourhood $V_p$ and $S_p=\partial V_p$, where we assume that the surface $S_p$ contains no singular points.   The space of unit length vectors in $D$ dimensions is isomorphic to the $D-1$ dimensional unit length sphere $S^{(D-1)}$, so the vector field $v$ defines a mapping:
\[
S_p\rightarrow S^{(D-1)}
\]
Because by assumption $S_p$ contains no singular points, it is homotopic to $S^{(D-1)}$, hence the above mapping is classified by the homotopy group $\pi_{(D-1)}(S^{(D-1)})\equiv \mathbb{Z}$.  The meaning of this is that the quantity $\mu$ takes integer values and it is invariant to deformations in the region $S_p$ that do not cross singularities; i.e. it is a topological invariant.

If we choose the neighbourhood $V_p$ small enough then the only singularity will be $p$ and the index characterising it is
\[
\mu_p=\frac{\Gamma(D/2)}{2\pi^{D/2}}\oint_{S_p} \omega\in\mathbb{Z}-\{0\}.
\]

\section{Proof for the position of the centroids when the system has N-fold rotational symmetry}\label{app_proof}
If the weights are solely a function of the intensity value, then an intensity pattern with rotational symmetry (cylindrical or discrete rotational symmetry) will yield a point at the symmetry axis. Defining the symmetry axis as the origin, this means $\vec{p}_{\mathrm{i}}=\vec{0}$. We will show this for N-fold rotational symmetry. In order to simplify the notation we will drop the dependence of the intensity $I(y_1,y_2;x_1,x_2)$ on the position of the object $(x_1,x_2)$ and will drop the subindex of the vector $\vec{p}_i$, as the proof is the same for all vectors.

\begin{eqnarray}
\vec{p}&=&\int \int \mathrm{h}\left(I\left(y_1,y_2\right)\right)\left(y_1\hat{y}_1+y_2\hat{y}_2\right)~\mathrm{d}y_1~ \mathrm{d}y_2\nonumber \\
&=&\int_0^{2\pi}\int_0^{\infty}\mathrm{h}\left(I\left(r\cos\left(\theta \right),r\sin\left(\theta \right)\right)\right)\left(r\cos\left(\theta \right)\hat{y}_1+r\sin\left(\theta \right)\hat{y}_2\right)r\mathrm{d}r~ \mathrm{d}\theta \nonumber \\
&=&\sum\limits^{N-1}_{n=0} \int_0^{\frac{2\pi}{N}}\int_0^{\infty}\mathrm{h}\left(I\left(r\cos\left(\theta+\frac{2\pi n}{N} \right),r\sin\left(\theta+\frac{2\pi n}{N} \right)\right)\right)\nonumber \\
&&\times\left(r\cos\left(\theta+\frac{2\pi n}{N} \right)\hat{y}_1+r\sin\left(\theta+\frac{2\pi n}{N} \right)\hat{y}_2\right)r\mathrm{d}r~ \mathrm{d}\theta \nonumber \\
&=&\sum\limits^{N-1}_{n=0} \int_0^{\frac{2\pi}{N}}\int_0^{\infty}\mathrm{h}\left(I\left(r\cos\left(\theta \right),r\sin\left(\theta \right)\right)\right)\nonumber \\
&&\times\left(r\cos\left(\theta+\frac{2\pi n}{N} \right)\hat{y}_1+r\sin\left(\theta+\frac{2\pi n}{N} \right)\hat{y}_2\right)r\mathrm{d}r~ \mathrm{d}\theta \nonumber \\
&&\nonumber \\
&=&\vec{0}.
\end{eqnarray} 
In the second equality we have performed a change of coordinates: $(y_1,y_2)=(r \cos(\theta), r\sin(\theta))$. In the third equality we have divided the azimuthal integral in $N$ different parts. In the fourth equality we have used the assumed $N-$fold rotational symmetry of the intensity distribution. The last equality is just a result of the summation over the $\sin$ and $\cos$ functions.

\section*{References}
\bibliography{References}

\end{document}